\documentclass[12pt]{article}
\def\bea{\begin{eqnarray}}
\def\eea{\end{eqnarray}}
\usepackage{graphicx}

\begin{document}

\begin{center}
{\large\bf The level crossing and inverse statistic analysis of
German stock market index (DAX) and daily oil price time series }

\bigskip
F. Shayeganfar,$^a$ M. H$\ddot{o}$lling,$^b$  J. Peinke,$^b$  and M.
Reza Rahimi Tabar$^{a,c}$

{\it $^a$Department of Physics, Sharif University of Technology,
P.O.Box 11365-9161, Tehran 11365, Iran\\
$^b$Carl von Ossietzky University, Institute of Physics, D-26111
Oldenburg, Germany\\
$^c$ Fachbereich Physik, Universit\"{a}t Osnabr\"{u}ck,
Barbarastra{\ss}e 7, 49076 Osnabr\"{u}ck, Germany }

\end{center}

The level crossing and inverse statistics analysis of DAX and oil
price time series are given. We determine the average frequency of
positive-slope crossings, $\nu_{\alpha}^+$, where $T_{\alpha}
=1/\nu_{\alpha}^+ $ is the average waiting time for observing the
level $\alpha$ again.
 We estimate the probability $P(K, \alpha)$, which provides us the probability of
observing $K$ times of the level $\alpha$ with positive slope, in
time scale $T_{\alpha}$. For analyzed time series we found that
maximum $K$ is about $\approx 6$. We show that by using the level
crossing analysis one can estimate how the DAX and oil time series
will develop. We carry out same analysis for the increments of DAX
and oil price log-returns,(which is known as inverse statistics)
 and provide the distribution of waiting times to observe some level for the increments. \\


\newpage

\section{Introduction}

Stochastic processes occur in many natural and man-made phenomena,
ranging from various indicators of economic activities in the stock
market, velocity fluctuations in turbulent flows and heartbeat
dynamics, etc \cite{1}. The level crossing analysis of stochastic
processes has been introduced by (Rice, 1944, 1945) [2-26], and used
to describe the turbulence \cite{2-14}, rough surfaces \cite{3},
stock markets \cite{4}, Burgers turbulence and Kardar-Parisi-Zhang
equation \cite{5,6}. The level crossing analysis of the data set has
the advantage that it gives important global properties of the time
series and do not need the scaling feature. The almost of the
methods in time series analysis are using the scaling features of
time series, and their applications are restricted to the time
series with scaling properties. Our goal with the level crossing
analysis is to characterize the statistical properties of the data
set with the hope to better understand the underlying stochastic
dynamics and provide a possible tool to estimate its dynamics. The
level crossing and inverse statistics analysis can be viewed as the
complementary method to the other well-known methods such as,
detrended fluctuation analysis (DFA) \cite{7}, detrended moving
average (DMA) \cite{8}, wavelet transform modulus maxima (WTMM)
\cite{9}, rescaled range analysis (R/S) \cite{10}, scaled windowed
variance (SWV) \cite{11}, Langevin dynamics \cite{12}, detrended
cross-correlation analysis \cite{13}, multifactor analysis of
multiscaling \cite{14}, etc.

We start with formalism of the level crossing analysis. Consider a
time series of length $n$ given by ${x(t_{1}), x(t_{2}),...,
x(t_{n})}$ (here $x(t_i)$ is the log-return of DAX and oil prices).
The log-return $x(t_i)$ is defined as $x(t_i)= \ln (y_i / y_{i-1})$,
where $y_i$ is the price at time $t_i$. Let $N_{\alpha}^{+}$ denote
the averaged number of $positive$ slope crossing of $x(t)= \alpha$
in time scale $T=n \Delta t$ with $\Delta t=1$ (we set also the
average $< x
> $ to be zero). The averaged $N_{\alpha}^{+}$ can be written as  $
N_{\alpha}^{+}(T)=\nu^{+}_{\alpha} T $, where $\nu_{\alpha}^{+}$ is
the average frequency of positive slope crossing of the level
$\alpha$. The positive level crossing has specific importance that
it gives the next average time scale that the price $y_i$ will be
greater than the $y_{i-1}$ again up to specific level. For narrow
band processes it has been shown that the frequency
$\nu_{\alpha}^{+}$ can be deduced from the underlying joint
probability distributions function (PDF) for $x$ and $dx/dt=
\dot{x}$. Rice proved that \cite{2}
\begin{equation}
\nu_{\alpha}^{+}= \int_0 ^{\infty} \dot{x} p(x=\alpha,\dot{x})
d\dot{x},
\end{equation}
where $p(x,\dot{x})$ is the joint PDF of $x$ and $\dot{x}$.
 For discrete time series (of course all of real data are discrete) the
frequency $\nu_{\alpha}^{+}$ can be written in terms of joint
cumulative probability distribution, $P(x_i>\alpha , x_{i-1} <
\alpha)$ as \cite{15},
\begin{eqnarray}
\nu_{\alpha}^+  &=& P(x_{i} >\alpha , x_{i-1} < \alpha)\cr \nonumber
\\ &=&
 \int_{-\infty}^{\alpha} \int_{\alpha}^{\infty}
p(x_i,x_{i-1})dx_i dx_{i-1},
\end{eqnarray}
where $p(x_i,x_{i-1})$ is the joint PDF of $x_i$ and $x_{i-1}$. The
inverse of frequency $\nu_{\alpha}^{+}$ gives the average time scale
$T_{\alpha}$ that one should wait to observe the given level
$\alpha$ again.

The rest of this paper is organized as follows: Section 2 is devoted
to summary of level cross analyzing of DAX and daily oil price
log-returns. The inverse statistics of DAX and Oil price time series
are given in section 3. Section 4 closes with a discussion and
conclusion of the present results.

\section{level crossing}

Here, at first we provide the results of level crossing analysis for
two normalized log-return time series, daily German stock market
index (the DAX) and daily oil price. The daily fluctuations in the
oil price and DAX time series were belong to the period 1998-2009.
We also study the asymmetric properties of level crossing analysis
for positive and negative level crossing of the time series. To have
a comparison we provide also level crossing analysis of synthesized
uncorrelated noise. Also we will provide the results of level
crossing analysis of high frequency data for DAX with sample rate
$4$ (1/min), where we have used $2511000$ data points and belongs to
the period 1994-2003.

Figure 1 shows the frequency $\nu_{\alpha}^{+}$ for daily
log-returns of DAX and synthesized uncorrelated noise. The PDF of
synthesized uncorrelated noise is Gaussian and it has white noise
nature (i.e. its correlation has delta-function behavior). As shown
in figure 1, their level crossing frequency are almost similar near
to $\alpha \simeq 0$ and have deviation for levels in the tails. The
difference is related to the non-Gaussian PDF of DAX log-return time
series (see below). For the normalized Gaussian uncorrelated noise
one can show that the frequency $\nu_{\alpha}^{+}$ is given by
\begin{equation}
\nu_{\alpha}^{+}=  \frac{1}{4} [ 1- erf^2(\alpha/\sqrt{2})],
\end{equation}
where $erf(U)$ is the error function. In figure 1 a comparison
between the analytical and numerical results for uncorrelated
Gaussian time series is given. For Gaussian uncorrelated data the
frequency $\nu_{\alpha}^{+}$ behaves as: \bea
\nu_{\alpha}^{+} &\simeq&  \frac{1}{4} \exp(- \alpha^2 /2 \pi) \hskip 0.5cm  for \hskip 0.5cm \alpha  \to 0  \cr \nonumber \\
\nu_{\alpha}^{+} &\simeq&   \frac{1}{4} \exp(- 4 \alpha^2 / \pi)
\hskip 0.5cm for  \hskip 0.5cm \alpha  \to \pm \infty, \eea
 while we have found that $\nu_{\alpha}^{+}$ for daily DAX and oil price time series have
 power-law tails $ 1/ |\alpha|^{\beta} $ with exponents $ \beta_{DAX}
 = 3.5 \pm 0.1$ and $ \beta_{oil}
 = 3.8 \pm 0.2$, respectively (see inset of figure 1). It means that
 the DAX and oil price time series have non-Gaussian tails for their level crossing.
We note that within the error bars, the exponent $\beta_{DAX} \simeq
\beta_{oil}$. However the exponent $\beta$ may depends on the sample
rate of data acquisition (see below). The exponent $\beta$ can be
estimated using the method proposed in \cite{20,20a,20b}. If the PDF
or $\nu_{\alpha}^{+}$ follow a power law with exponent $\beta=\kappa
+1$, one can estimate the power-law exponent $\kappa$ by sorting the
normalized returns or levels by their sizes, $ \alpha_1 > \alpha_2
> ... > \alpha_ N$, with the result \cite{20a} $ \kappa = (N - 1) [
\sum _{i=1} ^ {N-1}  \ln \frac {\alpha_i}{ \alpha_N}]^{-1}$, where
(N - 1) is the number of tail data points.

In general, there are two reasons to have non-Gaussian tails for
level crossing of given time series; (i) due to the fatness of the
probability density function (PDF) of the time series, in comparison
to a Gaussian PDF. By definition a fat PDF is defined via the
behavior of its tails. If its tail goes to the zero slower than a
Gaussian PDF then we call it as fat tail PDF.
 In this case, non-Gaussian tails cannot be
changed by shuffling the series, because the correlations in the
data set are affected by the shuffling, while the PDF of the series
is invariant, (ii) due to the long-range correlation in time series.
In this case, the data may have a PDF with finite moments, e.g., a
Gaussian  distribution. The easiest way to distinguish whether the
PDF shape or long-range correlation is responsible for the
fastnesses of $\nu_{\alpha} ^+$ for the DAX and oil log-returns time
series, is by analyzing the corresponding shuffled and surrogate
time series. The level crossing analysis will be sensitive to
correlation when the time series is shuffled and to probability
density functions (PDF) with fat tails when the time series is
surrogated. The long range correlations are destroyed by the
shuffling procedure and in the surrogate method the phase of the
discrete Fourier transform coefficients of time series are replaced
with a set of pseudo-independent distributed uniform $(-\pi,+\pi)$
quantities. The correlations in the surrogate series do not change,
but the probability function changes to Gaussian distribution
\cite{16,17,17-1}.

In figure 1 the level crossing frequency of shuffled and surrogated
DAX time series are given. The figure shows that the frequency
$\nu_{\alpha}^{+}$ has more difference for original and surrogated
time series and means that the non-Gaussian tails for
$\nu_{\alpha}^{+}$ due to the fatness of PDF is dominant
\cite{2-14}. We have found similar results for daily oil price
log-returns.

Now let us introduce the PDF, $P(K,\alpha)$, which provides us the
probability of observing $K$ times of the level $\alpha$ with
positive slope, in the averaged time scale $T_{\alpha}$. By
construction the average $< K >$, i.e.  $< K >|_{\alpha} =
\sum_{K=0} ^{N} K P(K, \alpha)$ will be unity and $P(K,\alpha)$
satisfies the normalization condition $\sum_{K=0} ^{N} P(K, \alpha)
= 1$. In principle we can assume that the upper bound, i.e. $N$ to
be infinity. For the processes and levels that satisfy $ P(0,\alpha)
<< \sum_{K=1} ^ N P(K,\alpha) $,  we expect to have a good
estimation about the future of process. This means that one will
observe the level $\alpha$ with high probability in time scale
$T_{\alpha}$ at least once. For the levels that the PDF
$P(K,\alpha)$ satisfies $\sum_{K=1} ^ N P(K,\alpha) \approx
P(0,\alpha)$, the process will be not predictable. Figure 2 shows
the PDF $P(K,\alpha)$, for daily DAX and oil price log-return time
series for different levels $\alpha$. We plotted also $P(K,\alpha)$
for some levels of synthesized Gaussian uncorrelated data to have a
comparison. As shown in figure 2 the upper bound $N$ is about 6,
which means that it is almost impossible to observe same level
$\alpha$ in average time scale $T_{\alpha}$ more than 6 times, even
for the white noise. We used $10^7$ data points for white noise
synthesized data and found that the maximum number of observing is
also about 6.

To find the best interval for the estimation of time series future,
we consider the variation of the PDF, $P(0,\alpha)$ with respect to
the level $\alpha$. This will enable us to find the range and
intervals of levels that one can estimate the future of these time
series with high accuracy. In figure 3 the PDFs $ P( K=0 , \alpha )$
for daily DAX, oil and uncorrelated synthesized time series, are
given. It shows that the $ P( K=0 , \alpha )$ of daily DAX  and oil
time series for the interval $ -0.5 < \alpha < 0.5 $ has smaller
probability with respect to uncorrelated time series. It means that
with high probability (with respect to white noise), one can observe
the level $\alpha$ at least once in time scale $T_{\alpha}$ for
$\alpha$s belong to this interval. The typical time scale
$T_{\alpha}$ for these interval is about 4 days for the daily DAX
and oil time series, respectively. The corresponding time scales
$T_{\alpha}$ for different $\alpha$s are shown in figure 4. For the
daily DAX and oil price time series (for the interval $  2 > \alpha
> -2 $) we found the following empirical curve
fittings: \bea
T_{\alpha} (DAX) &=& 4.10 - 0.18 \alpha + 4.90 \alpha^2+0.21 \alpha^3 + 0.94 \alpha^4 ,\cr \nonumber \\
T_{\alpha} (oil) &=&  3.90 - 0.37 \alpha + 5.70 \alpha^2 + 0.57
\alpha^3+ 1.69 \alpha^4 \eea where $T_{\alpha}$`s have the dimension
"days". To check the applicability level crossing for forecast of
the time series, in figure 5 we plotted the daily time series of DAX
and indicates the points with level $\alpha=0$ with red points. The
average $T_\alpha$ for this level indicated by vertical lines.
 We expect that in this time scale one should observe another red
 points with high probability. We also investigated the asymmetry properties of time series with
respect to positive and negative slops.

Finally, we have done similar analysis to the high frequency DAX
time series and find that the best interval to estimate the future
is $ -0.01 < \alpha <  0.01 $. The typical time scale belong to this
levels is about $75$sec and the obtained exponent $\beta$ was $ 2.4
\pm 0.1$. For this time series, the averaged time scale $T_{\alpha}$
depends on the level $\alpha$ as:

\bea T_{\alpha} (DAX) = 72 - 3 \alpha+ 186 \alpha^2 +  \alpha^3 - 4
\alpha^4, \eea where $T_{\alpha}$ has dimension in seconds.

\section{Inverse statistics}

To the modeling the statistical properties of financial time series
Simonsen, et al. \cite{42} asked the "inverse" question: what is the
smallest time interval needed for an asset to cross a fixed return
level $\gamma$? or what is the typical time span needed to generate
a fluctuation or a movement in the price of a given size [42-46]?
The inverse statistics is the distribution of waiting times needed
to achieve a predefined level of return obtained from every time
series. This distribution typically goes through a maximum at a time
so called the optimal investment horizon, which is the most likely
waiting time for obtaining a given return \cite{47}. Let ${y(t)}$ be
the price at time $t$. The logarithmic return calculated over the
interval $\Delta t$ is, $r_{\Delta t}(t)= \ln(y(t+\Delta t)) -
\ln(y(t))$. Given a fixed log-returned barrier, $\gamma$, of an
index, the corresponding time span is estimated for which the
log-return of index for the first time reaches the level $\gamma$.
This can also be called the first passage time through the level
$\gamma$ for $r_{\Delta t}$.  In figure 6, we plotted the
probability distribution $p(\tau)$ of normalized waiting time $\tau$
needed to reach return levels $\gamma = 0, 1\sigma, 2\sigma$ for
daily oil, daily DAX log-returns and integrated white noise data
(i.e. fractional Brownian motion fBm).

As figures 6 show, for the zero level for $r_{\Delta t}$, inverse
 statistics of the two markets does not deviate from fractional
 Brownian motion while they are rather different behavior from fBm
 for $\gamma = 1\sigma$ and $2\sigma$. We fit the waiting times
 distribution functions $p(\tau)$ for different level $\gamma
$ via the Weibull distribution function \cite{21}:
\begin{eqnarray}
p(\tau,T)=\frac{\delta}{T}(\frac{\tau}{T})^{\delta -
1}\exp[-(\frac{\tau}{T})^\delta]
\end{eqnarray}
Where $\delta$ is the stretched exponent (or shape parameter) and
$T$ is the characteristic time scale. We found $\delta$ and $T$ for
fBm and Oil and DAX time series and summarized results in Table 1.

 \begin{center}
TABLE I. The stretched exponent $\delta$ and characteristic time
scale $T$ fitted by Weibull distribution for various time series.

\bigskip
\begin{tabular}{cccc}
\hline\hline
${\gamma}$ & ${time series}$ & ${\delta}$ & ${T}$\\
\hline
0  & fBm  & 0.423 & 1.849\\
\hline
0  & Oil  & 0.347 & 1.132\\
\hline
0  & DAX  & 0.312 & 5.791\\
\hline
1  & fBm  & 0.704 & 13.145\\
\hline
1  & Oil  & 0.496 & 7.943\\
\hline
1  & DAX  & 0.600 & 6.836\\
\hline
2  & fBm  & 0.956 & 34.178 \\
\hline
2  & Oil  & 0.878 & 32.480\\
\hline
2  & DAX  & 0.988 & 18.631\\
\hline\hline
\end{tabular}
\end{center}

\section{Conclusion}

In summary, we analyzed the DAX and oil daily price log-return time
series using the level crossing method and find the average waiting
time $T_{\alpha}$ for observing the level $\alpha$ again. This is a
similar analysis as what has been done in Refs. \cite{18,19,20}.
They have been carried out the level crossing of the volatility time
series, instead of the time series itself. We define and estimate
the probability of observing K times of the level $\alpha$, $P(K,
\alpha)$ in time scale $T_{\alpha}$.
  We show that by using the level
crossing analysis one can estimate the future of the daily DAX and
oil time series with good precision for the levels in the interval
$-0.5 < \alpha < 0.5$. Also, using the inverse statistics we
estimate the waiting time probability distribution for two financial
markets, i.e. oil and DAX time series.

\newpage

\newcounter{bean}
\begin{list}%
{[\arabic{bean}]}{\usecounter{bean}\setlength{\rightmargin}{\leftmargin}}

\bibitem{1} R. Mantegna and H.E. Stanley, {\it An Introduction to
Econophysics: Correlations and Complexities in Finance} (Cambridge
University Press, New York, 2000);  R. Friedrich, J. Peinke and M.
Reza Rahimi Tabar, "Importance of Fluctuations: Complexity in the
View of Stochastic Processes" Encyclopedia of Complexity and Systems
Science, 3574 (Springer, Berlin 2009).
\bibitem{2} S. O. Rice,  Bell System Tech. J. {\bf 23} 282 (1944); ibid. {\bf
24} 46 (1945).
\bibitem{2-1}W. Liepmann, Helv. Phys. Acta {\bf 22}, 119 (1949); H. W. Liepmann
and M. S. Robinson, NACA TN, 3037 (1952).
\bibitem{2-2} H. Steinberg, P.M. Schultheiss, C. A.Wogrin and F. Zwieg, J. Appl. Phys. {\bf 26} 195 (1955).
\bibitem{2-3} J. A. McFadden, IEEE Trans. Inform. Theory IT-{\bf 2} 146 (1956); IT-{\bf 4} 14 (1958).
\bibitem{2-4} C. W. Helstrom, IEEE Trans. Inform. Theory IT-{\bf 3} 232
(1957).
\bibitem{2-5} I. Miller and J. E. Freund, J. Appl. Phys. {\bf 27} 1290
(1958).
\bibitem{2-6} A. J. Rainal, IEEE
Trans. Inform. Theory IT-{\bf 8}, 372 (1962).
\bibitem{2-7} D. Slepian, Proc. Sympos. Time
Series Analysis, Brown Univ., ed. M. Rosenblatt, Wiley, 1963, pp.
104—115.
\bibitem{2-8} K. Ito, J. Math. Kyoto Univ. 3 (1963/1964) 207—216.
\bibitem{2-9} S. M. Cobb, IEEE Trans. Inform.
Theory IT-{\bf 11} 220 (1965).
\bibitem{2-10} N. D. Ylvisarer, Annals of Math. Statist. {\bf 36} 1043 (1965).
\bibitem{2-11} M. R. Leadbetter and J. D. Cryer, Bull. Amer. Math. Soc. {\bf
71} 561 (1965); R. N. Miroshin, St. Petersburg Univ. Math. {\bf 34}
30 (2001).
\bibitem{2-12} Orey, Z. Wahrscheinlichkeitstheor. Verwandte Geb. {\bf 15}, 249
(1970)
\bibitem{2-13} J. Abrahams, IEEE Trans. Inform. Theory IT-{\bf 28} 677 (1982).
\bibitem{2-14} K. R. Sreenivasan, A. Prabhu, and R. Narasimha, J. Fluid Mech. {\bf
137}, 251 (1983); K. R. Sreenivasan, Annu. Rev. Fluid Mech. {\bf
23}, 539 (1991).
\bibitem{2-15} A. J. Rainal, IEEE Trans. Inform. Theory
{\bf 34} 1383 (1988); A. J. Rainal, IEEE Trans. Inform. Theory {\bf
36} 1179 (1990).
\bibitem{2-16} J. T. Barnett and B. Kedem, IEEE Trans. Inform. Theory {\bf 37} 1188 (1991);
G. L. Wise, ibib {bf 38} 213 (1992).
\bibitem{2-17} I. Rychlik, Extremes {\bf 3} 331 (2000).
\bibitem{2-18} J. Davila and J. C. Vassilicos, Phys. Rev. Lett. {\bf 91}, 144501
(2003).
\bibitem{2-19} St. L\"{u}ck, Ch. Renner, J. Peinke and  R.  Friedrich, Physics Letters A{\bf 359} 335 (2006).
\bibitem{2-20}  D. Hurst and J. C. Vassilicos, Phys. Fluids {\bf 19}, 035103 (2007).
\bibitem{2-21}  N. Mazellier and J. C. Vassilicos, Phys. Fluids {\bf 20}, 015101 (2008).
\bibitem{2-22} Poggi D, Katul, BOUNDARY-LAYER METEOROLOGY {\bf 136} 219
(2010).
\bibitem{2-23} M. F. Kratz and J. R. León, Extremes {\bf 13}
315 (2010).
\bibitem{2-24} N. Rimbert, Phys. Rev. E {\bf 81} 056315 (2010).
\bibitem{3} F. Shahbazi, S. Sobhanian, M. R. Rahimi Tabar, S. Khorram, G. R.
Frootan, and H. Zahed, J. Phys. A{\bf 36}, 2517 (2003).
\bibitem{4} G. R. Jafari, M. S. Movahed, S. M. Fazeli, M. R. Rahimi Tabar, and
S. F. Masoudi, J. Stat. Mech., P06008 (2006).
\bibitem{5} A. Bahraminasab, H. Rezazadeh, A. A. Masoudi, J. Phys. A: Math. Gen.
39, 3903–3909 (2006).
\bibitem{6} A. Bahraminasab M. S. Movahed , S. D. Nasiri , A. A. Masoudi, M.
Sahimi, Journal of Statistical Physics 124 (6): 1471-1490, (2006)
\bibitem{7} C.-K. Peng,  J. Mietus, J. M.  Hausdorff, S. Havlin, H. E.
Stanley, and  A. L. Goldberger,
 Phys. Rev. Lett.
{\bf70}, 1343(1993).
\bibitem{8} E. Alessio, A. Carbone,  G. Castelli, and V. Frappietro, Eur. Phys.
J. B {\bf27}, 197 (2002).
\bibitem{9} J. F. Muzy, E. Bacry, and A. Arneodo, Phys. Rev. Lett. {\bf 67},
3515 (1991).
\bibitem{10} H. E. Hurst, R. P. Black,  and  Y. M. Simaika, {\it Long-Term
Storage. An Experimental Study} (Constable, London, 1965).
\bibitem{11} A. Eke, P. Herman,  L.  Kocsis, and L. R. Kozak, Physiol. Meas.
{\bf 23}, R1-R38 (2002).
\bibitem{12} R. Friedrich and J. Peinke, Phys. Rev. Lett. {\bf 78}, 863 (1997);
G.R. Jafari, S.M. Fazlei, F. Ghasemi, S.M. Vaez Allaei, M. R. Rahimi
Tabar, A. Iraji Zad, and G. Kavei, Phys. Rev. Lett. {\bf 91}, 226101
(2003); F. Ghasemi,  J. Peinke, M. Sahimi and  M. Reza Rahimi Tabar,
Eur. Phys. J.  B  47, 411 (2005); P. Sangpour, G. R. Jafari, O.
Akhavan, A. Z. Moshfegh and M. Reza Rahimi Tabar, Phys. Rev.
B{\bf71} 155423 (2005); F. Ghasemi, M. Sahimi, J. Peinke and M. Reza
Rahimi Tabar, J. Biological Physics {\bf 32} 117 (2006); G. R.
Jafari,  M.  Sahimi,  M. Reza Rasaei  and M. Reza Rahimi Tabar,
Phys. Rev. E {\bf 83}, 026309 (2011); F. Shayeganfar, S.
Jabbari-Farouji, M. S. Movahed, G. R. Jafari  and M. Reza Rahimi
Tabar,  Phys. Rev. E {\bf 81}, 061404 (2010); A. Farahzadi, P.
Niyamakom, M. Beigmohammadi, N. Mayer, M. Heuken, F. Ghasemi, M.
Reza Rahimi Tabar, T. Michely and M. Wuttig, Europhysics Letters
{\bf 90}, 10008 (2010); S. Kimiagar, M. S. Movahed, S. Khorram and
M. Reza Rahimi Tabar, Journal of Statistical Physics {\bf 143}
148–167 (2011).

\bibitem{13} B. Podobnik and H. E. Stanley, Phys. Rev. Lett {\bf 100}, 084102
(2008).
\bibitem{14} F. Wang, K. Yamasaki, S. Havlin, and H. E. Stanley, Phys. Rev. E
{\bf 79}, 016103 (2009).
\bibitem{15} F. Ghasemi, M. Sahimi, J. Peinke,  R. Friedrich, G. R. Jafari, and
M. Reza Rahimi Tabar, Phys. Rev. E 75, 060102R  (2007)
\bibitem{16} J. Theiler and D. Prichard,  Fields Inst. Commun. {\bf 11},99
(1997).
\bibitem{17} T. Schreiber and A. Schmitz, Physica D {\bf 142}, 346 (2000).
\bibitem{17-1} B. Podobnik, D.F. Fu, H.E. Stanley and P.Ch. Ivanov, Eur. Phys. J. B {\bf 56}, 47
(2007).
\bibitem{42} I. Simonsen, M.H. Jensen, and A. Johansen, Eur.
Phys. J. B, {\bf 27} (2002).
\bibitem{43} M.H. Jensen, Phys. Rev. Lett {\bf 83}, 76 (1999).
\bibitem{44} L. Bifarele, M. Cencini, D. Vergni, A. Vulpiani, Eur.
Phys. J. B {bf 20}, 473 (2001).
\bibitem{45} S. Karlin, A First Course in Stochastic Processes (Academic Press, New York,
1966).
\bibitem{46} M. Ding, G.Rangarajan, Phys. Rev. E {\bf 52}, 207
(1995).
\bibitem{47} H. Ebadi, M. Bolgorian and G. R. Jafari, Physica A {\bf 389} 5530 (2010).
\bibitem{18} K. Yamasaki, L. Muchnik, S. Havlin, A. Bunde, and H. E. Stanley,
Proc. Natl. Acad. Sci. USA {\bf 102}, 9424 (2005).
\bibitem{19} F. Wang, K. Yamasaki, S. Havlin, and H. E. Stanley, Phys. Rev. E
{\bf 73}, 026117 (2006).
\bibitem{20} B. Podobnik, D. Horvatic, A. Peterson, H. E. Stanley,
 Proc. Natl. Acad. Sci. USA {\bf 106}, 22079 (2009).

\bibitem{20a}B. M. Hill, A simple general approach to inference about the
tail of a distribution. Ann Stat 3, 1163 (1975).

\bibitem{20b} A. Pagan, The econometrics of financial markets. J
Empirical Finance 3, 15 (1996).

\bibitem{21} P. C. Ivanov, A. Yuen, B. Podobnik and Y. Lee, Phys. Rev.
E {\bf 69}, 056107 (2004).
\end{list}%

\newpage

\noindent{\bf Captions}

FIG. 1. The level crossing analysis of the DAX log-returns
(original, shuffled and surrogated)
 and uncorrelated Gaussian  time series. Inset: the log-log plot of
 level crossing frequency vs level $\alpha$ for DAX log-return time
 series. The Gaussian  uncorrelated time series has exponential tails ($\sim \exp(- 4
 \alpha^2)$), while the daily DAX time series has power-low tails
 with exponent $\simeq 3.5$.

 FIG. 2. The PDf $P(K, \alpha)$ vs $K$ for different levels $\alpha$ for
 normalized log-returns time series of the daily German stock
market index (DAX) (top), oil daily price (bottom) and uncorrelated
synthesized data.

FIG. 3. The PDF $P(K=0,\alpha)$ vs $\alpha$ for DAX, oil daily price
and uncorrelated  synthesized normalized log-return time series. The
inset is same figure with wide range of $\alpha$`s. The results for
uncorrelated synthesized data are plotted to have a comparison.

FIG. 4. The level dependence of average time $T_{\alpha}$ for daily
DAX and oil price log-return time series.

FIG. 5.  The points (red) with level $\alpha=0$ for daily DAX time
series.  We expect that in this time scale one should observe
another red points with high probability.

FIG. 6. The probability distribution $p(\tau)$ of normalized waiting
time $\tau$ needed to reach return levels at scale $\tau$, i.e.
$\gamma = 0$, $\gamma = 1 \sigma$ and $\gamma = 2\sigma$ for two
financial markets including oil and DAX time series. Solid curves
are the fitted curve (Weibul distribution) based on Eq. (7).

\newpage

\begin{figure*}[htb]
\centerline{\includegraphics[scale=0.75,draft=false]{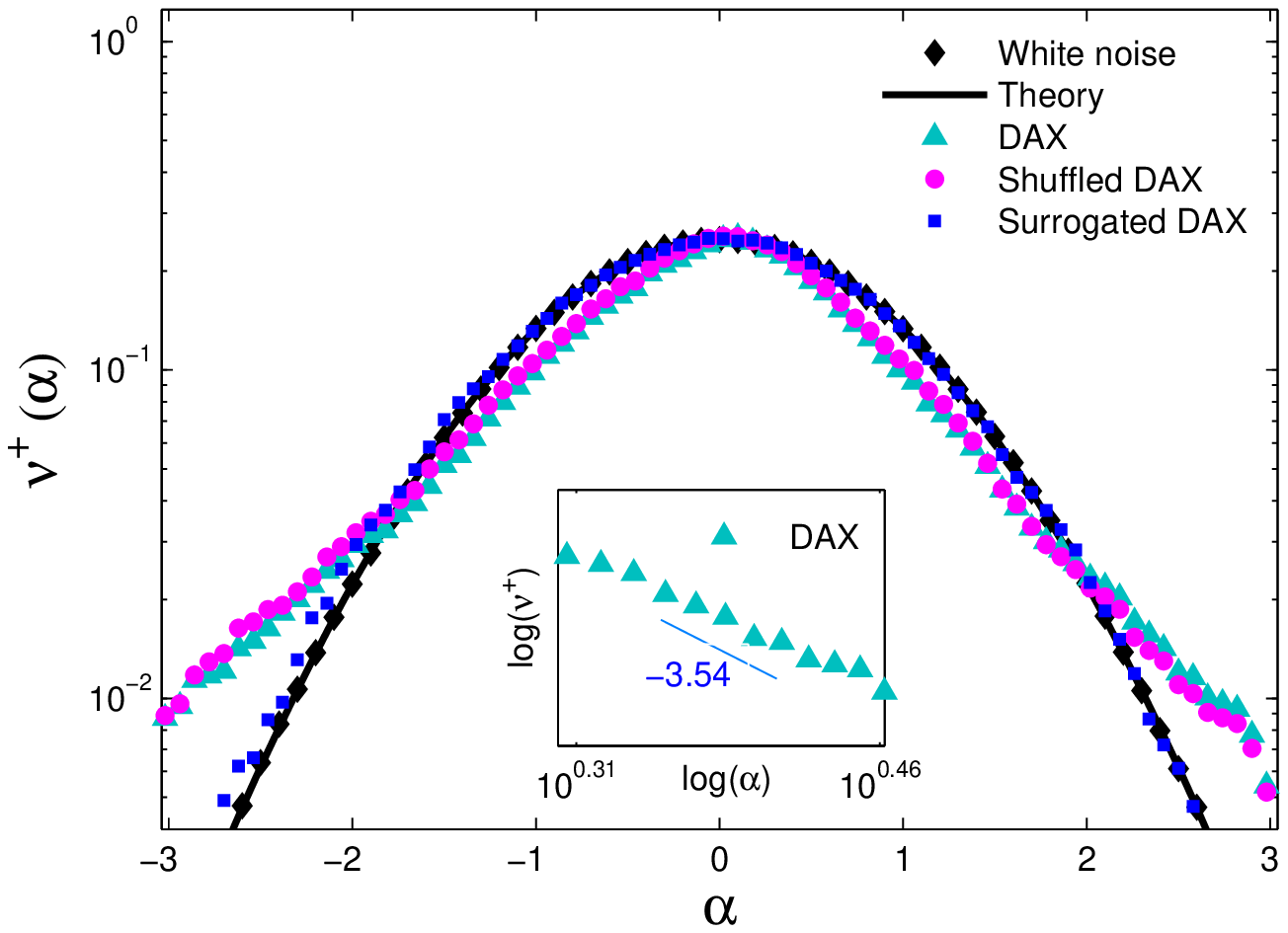}}
\caption{}
\end{figure*}

\newpage

\begin{figure*}[htb]
\centerline{\includegraphics[scale=0.45,draft=false]{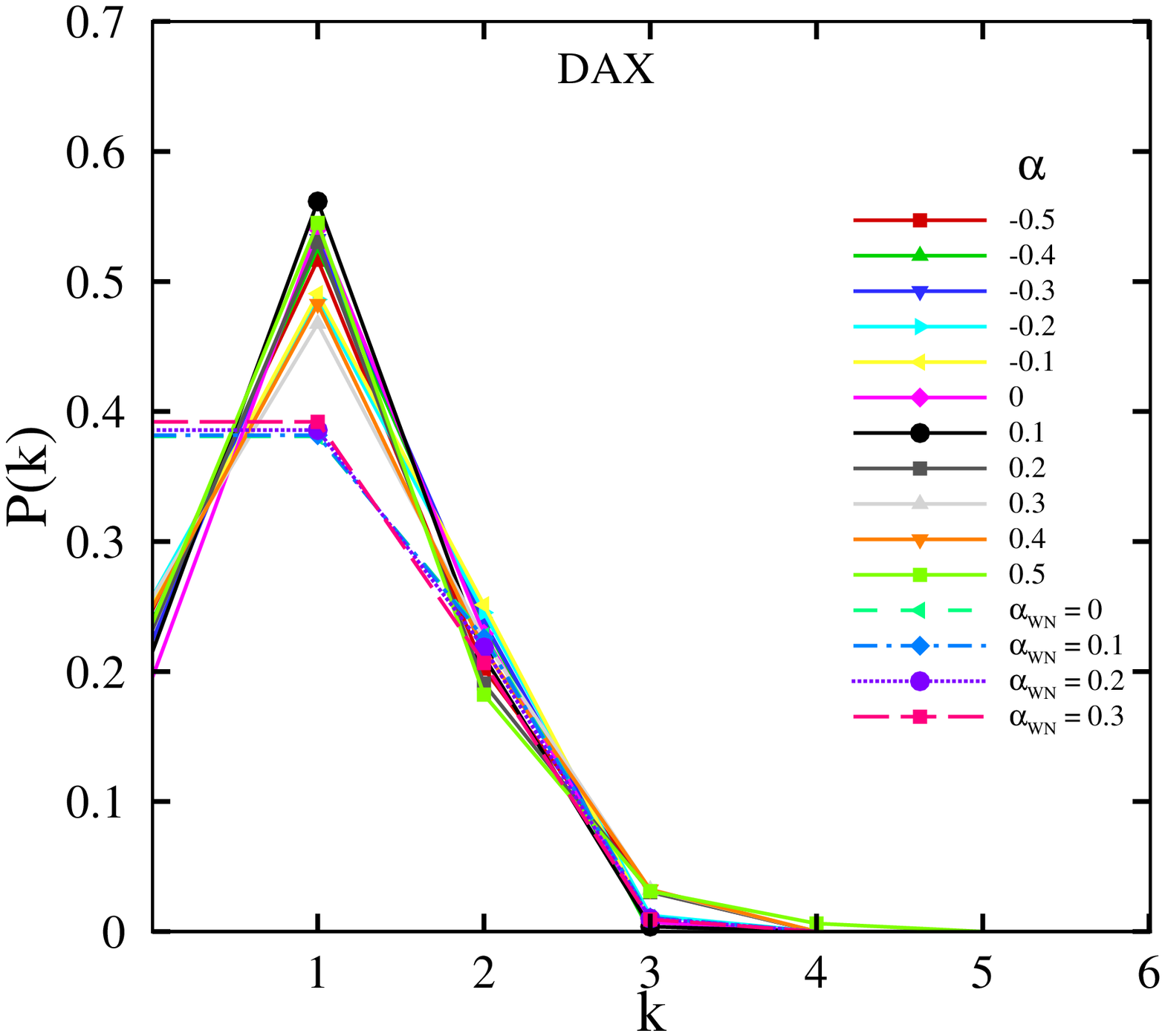}}
\centerline{\includegraphics[scale=0.45,draft=false]{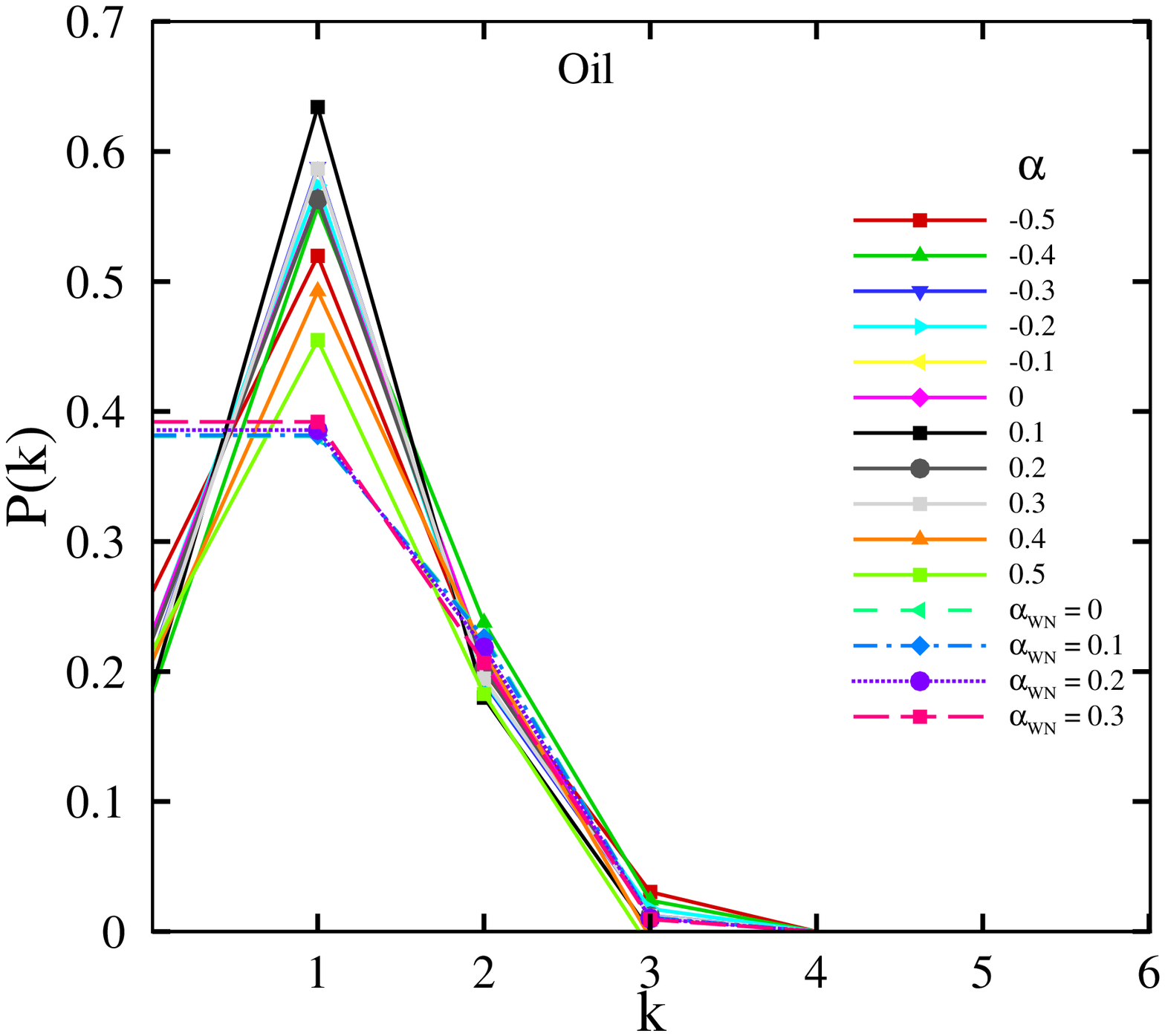}}
\caption{}
\end{figure*}

\newpage

\begin{figure*}[htb]
\centerline{\includegraphics[scale=0.7,draft=false]{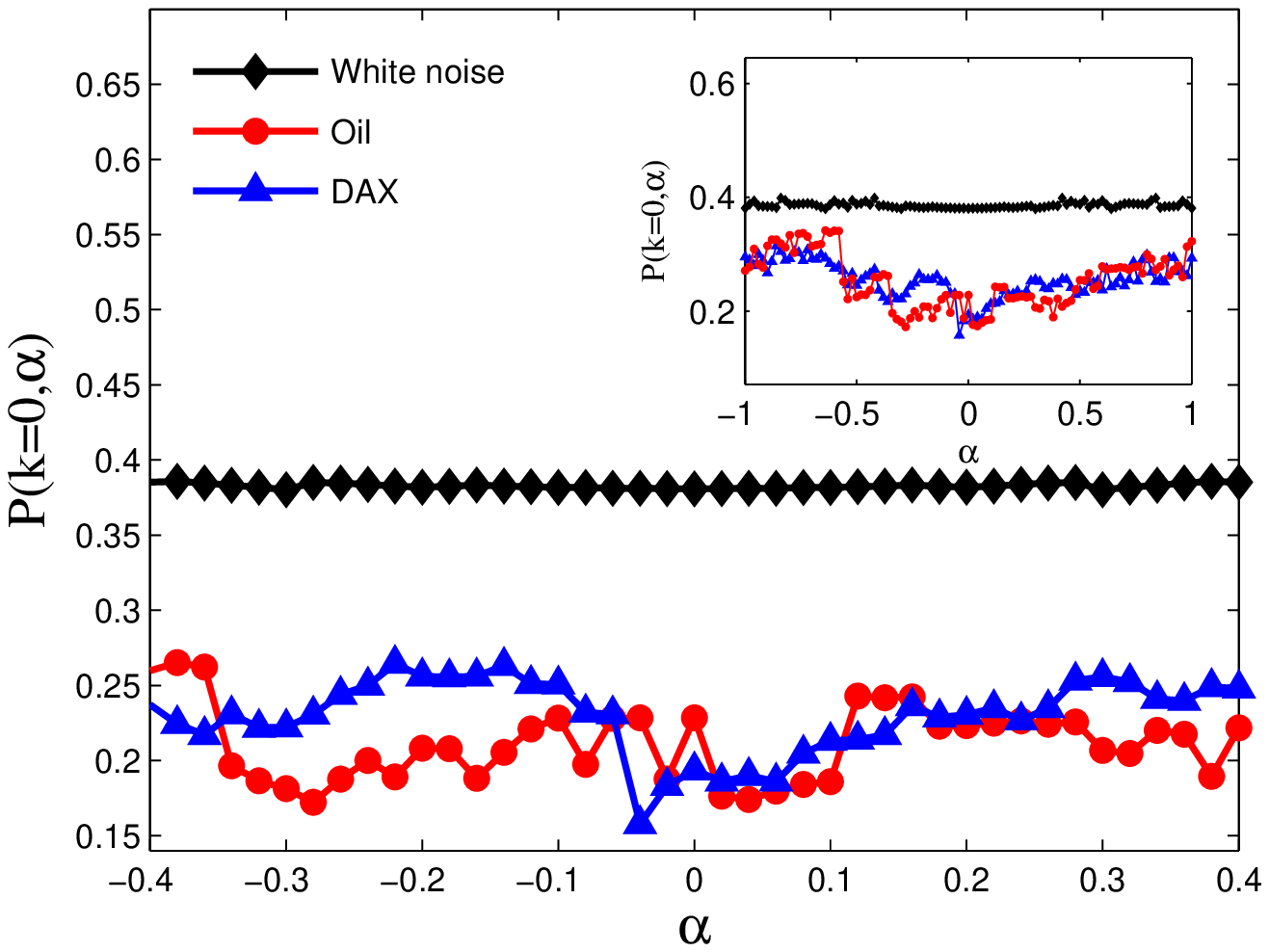}}
\caption{}
\end{figure*}
\newpage

\newpage

\begin{figure*}[htb]
\centerline{\includegraphics[scale=0.7,draft=false]{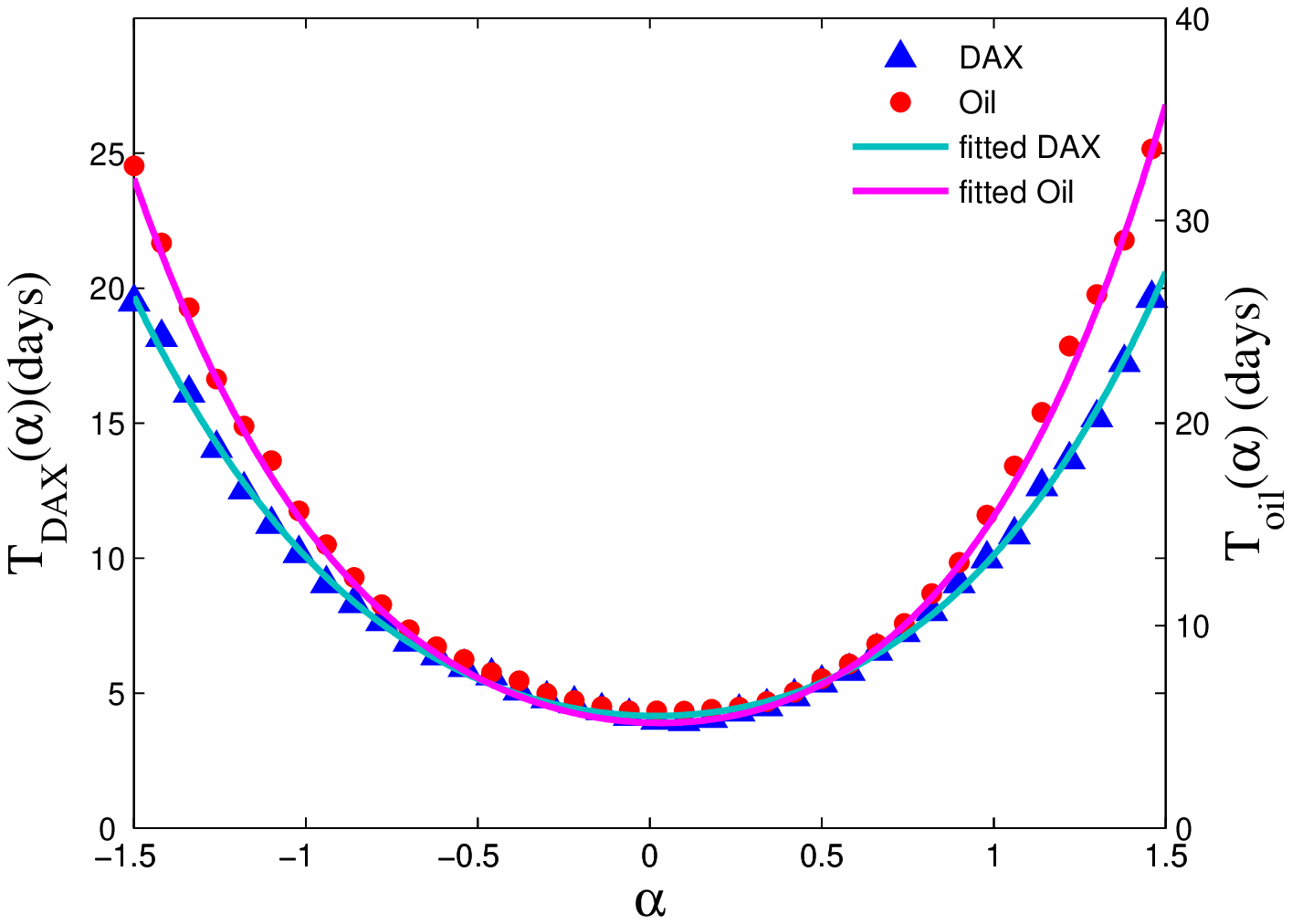}}
\caption{}
\end{figure*}

\newpage

\begin{figure*}[htb]
\centerline{\includegraphics[scale=0.8,draft=false]{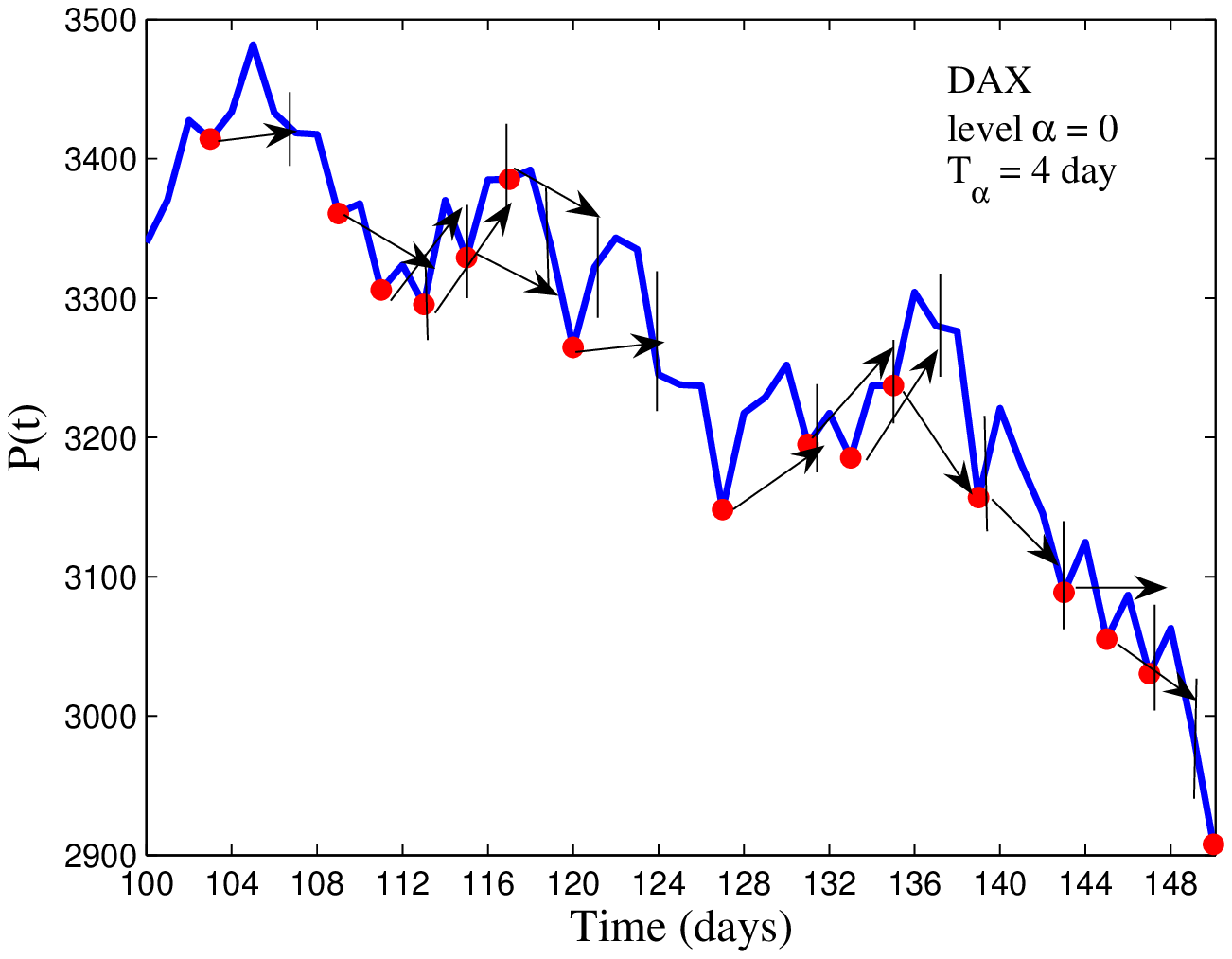}}
\caption{}
\end{figure*}

\newpage

\begin{figure*}[htb]
\centerline{\includegraphics[scale=0.35,draft=false]{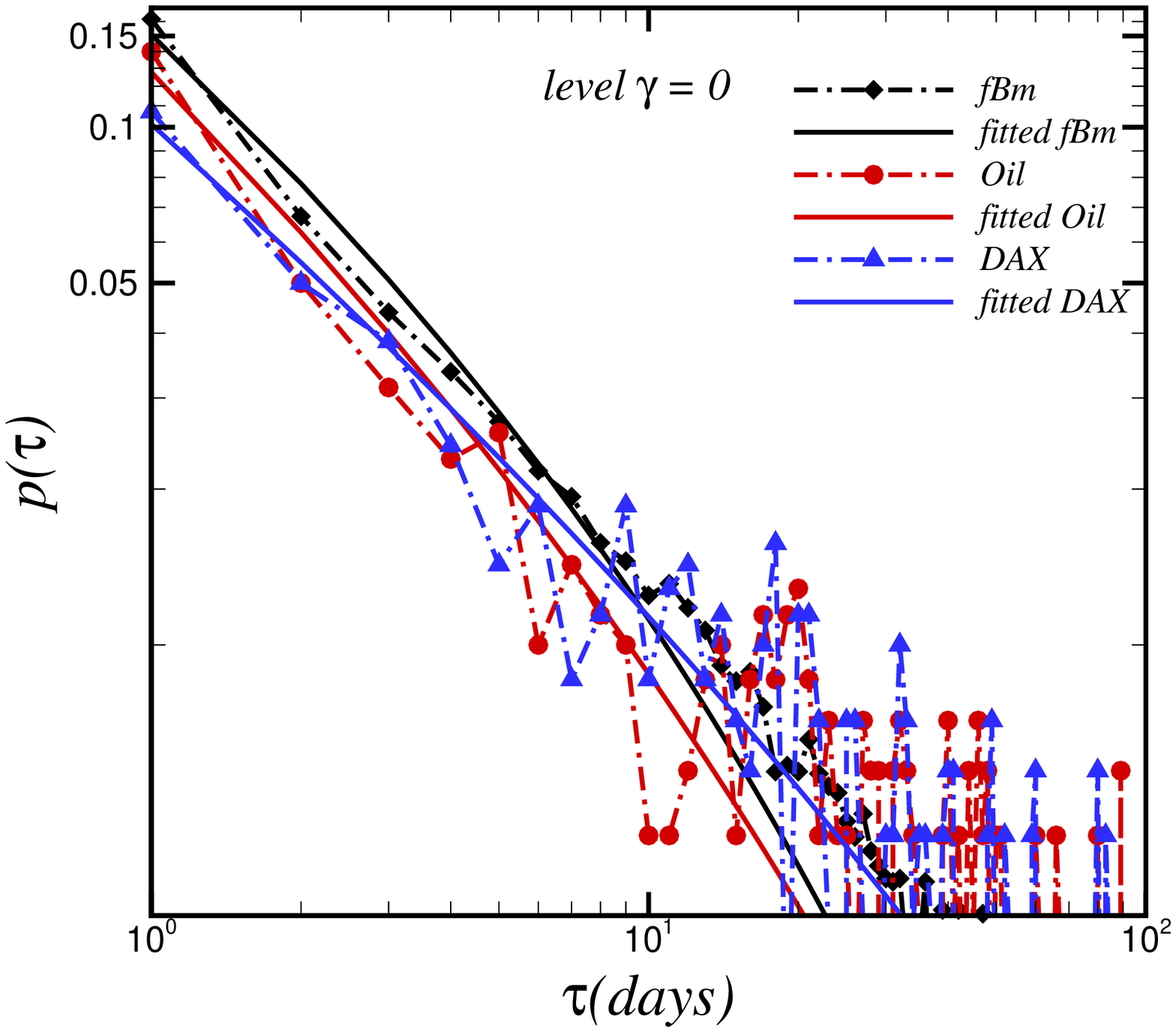}}
\centerline{\includegraphics[scale=0.35,draft=false]{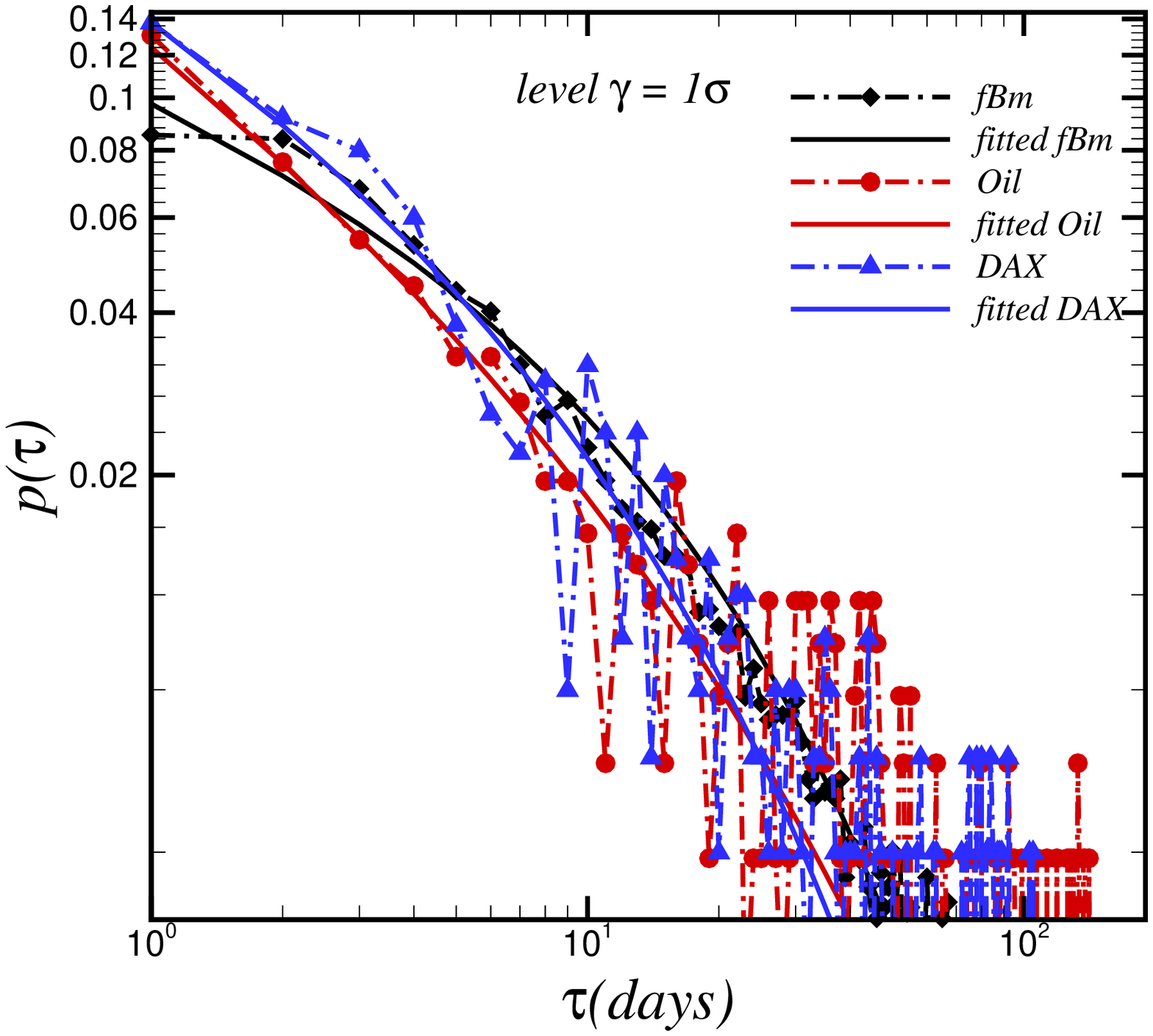}}
\centerline{\includegraphics[scale=0.35,draft=false]{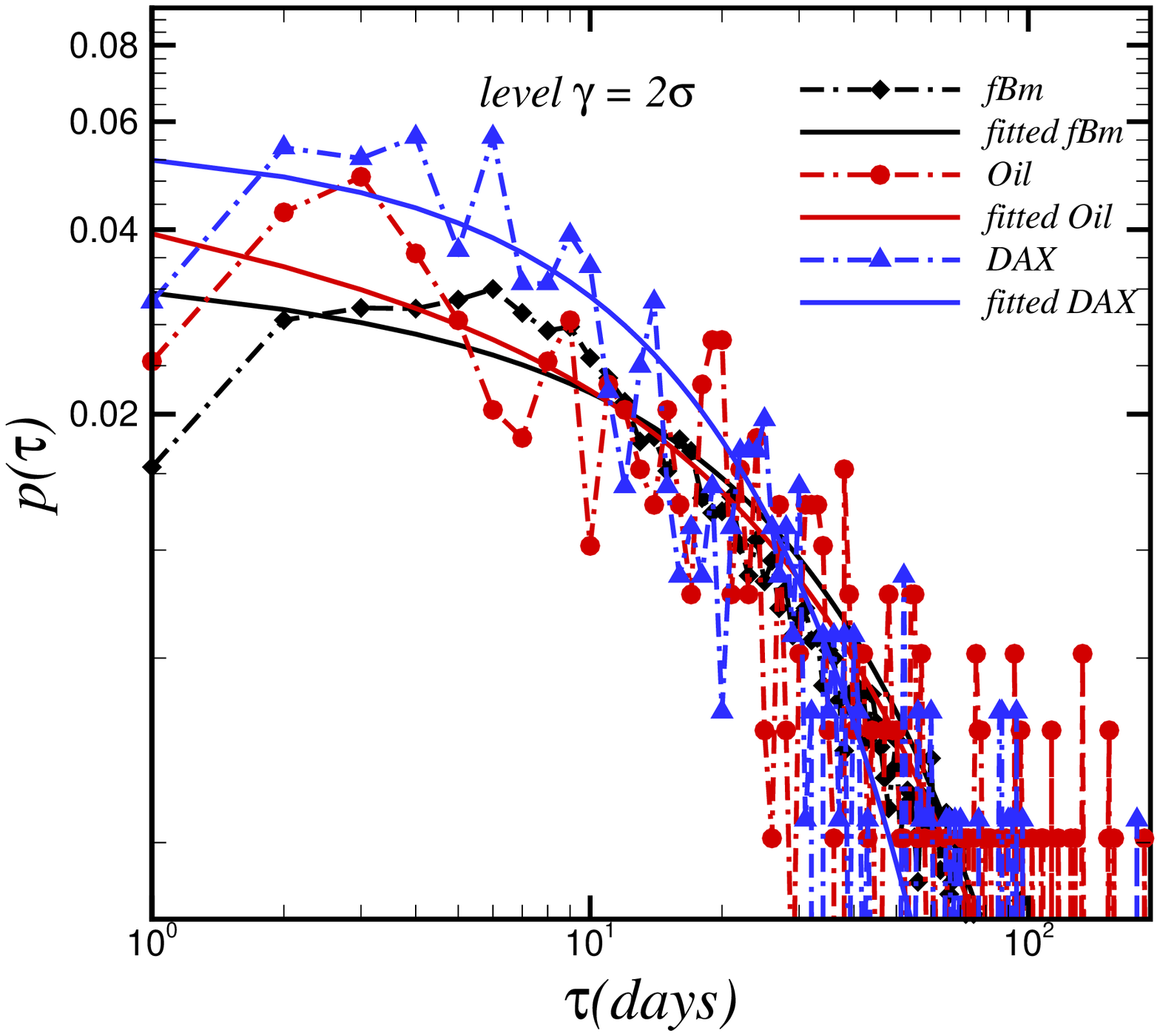}}
\caption{}
\end{figure*}

\end{document}